\documentclass[aps,prl,preprint,groupedaddress]{revtex4-1}

\usepackage{latexsym}
\usepackage{amsthm}
\usepackage{amsmath}
\usepackage{graphicx}
\usepackage{mathtools}
\usepackage{slashed}
\usepackage{amssymb}
\usepackage{indentfirst}
\usepackage{subfigure}
\usepackage{physics}
\usepackage{color}

\newcommand{\nn}{\nonumber}

\newcommand{\be}{\begin{eqnarray}}
\newcommand{\ee}{\end{eqnarray}}
\newcommand{\ma}{\mathrm}
\newcommand{\ml}{\mathcal}
\newcommand{\bs}{\boldsymbol}

\begin{document}

\title{An Effective Field Theory Approach to the Stabilization of $^8$Be in a QED Plasma}

\author{Xiaojun Yao$^1$, Thomas Mehen$^1$ and Berndt M\"uller$^{1,2}$}
\affiliation{$^1$Department of Physics, Duke University, Durham, NC 27708, USA\\
$^2$Brookhaven National Laboratory, Upton, NY 11973, USA}

\date{\today}

\begin{abstract}
We use effective field theory to study the $\ma{\alpha}$-$\ma{\alpha}$ resonant scattering in a finite-temperature 
QED plasma. The static plasma screening effect causes the resonance state $^8$Be to live longer and eventually leads to the formation of a bound state when $m_D\gtrsim 0.3$ MeV. We speculate that this effect may have implications on the rates of cosmologically and astrophysically relevant nuclear reactions involving $\ma{\alpha}$ particles.
\end{abstract}

\maketitle

Intuition suggests that plasma screening effects should be most visible in resonant nuclear reactions with a resonance located in the thermal energy domain. One example of such a reaction is the $^4\ma{He}(^4\ma{He},\gamma)^8\ma{Be}$ reaction (the $\alpha$-$\alpha$ reaction), which has been studied both experimentally and theoretically. A resonance is observed in the $s$-wave at the center-of-mass (CM) energy $E_{0}=91.84\pm0.04$ keV with a width $\Gamma_0=5.57\pm0.25$ eV, which corresponds to a lifetime of $1.2\times10^{-16}$ s \cite{Wustenbeckeretal.1992}. The resonance is identified as the ground state of $^8$Be with quantum numbers $(J^P, I)=( 0^+,0)$. The nuclear $s$-wave phase shift  has been measured up to $E_{CM}=10$ MeV \cite{Russell:1956zz}. Lying well inside the thermal domain (for example, of the $e^-e^+$ plasma in the early universe with $T\sim1$ MeV), the resonant $\alpha$-$\alpha$ reaction is an ideal candidate to study plasma screening effects on the position and width of the resonance. As we show below, the static screening prolongs the resonance lifetime and makes it possible to form a bound state.

The formalism of effective field theory (EFT) provides a framework to study the screening effects on low-energy nuclear scattering. As an example, we investigate in this letter the static screening effect on the resonant scattering of two $\alpha$ particles. The $\ma{\alpha}$ particle, $^{\ma{4}}\ma{He}$, is a tightly bound nucleus with charge $\ma{+2e}$ and isospin $I=0$. The internal nucleon dynamics has a momentum scale of the order of the pion mass $m_{\pi}\sim135$ MeV, which is much larger than the momentum at the resonance. Therefore, the internal structure of the $\alpha$ particle can be neglected in the low-energy resonance physics. As the $\ma{\alpha}$ particle mass $M\approx3727.38$ MeV is much larger than the energy scale of the scattering process, it can be treated non-relativistically. Furthermore, since one pion or one kaon exchange between two $\ma{\alpha}$ particles is forbidden by the isospin conservation, the exchange force can only be mediated by two pions or other heavier mesons with isospin $I=0$. The two-pion exchange sets an upper limit of the length scale of the $\ma{\alpha}$-$\ma{\alpha}$ interaction potential $\sim0.7\ \ma{fm}$. Note that the radius of the $\ma{\alpha}$ particle is roughly $2$ fm and the de Broglie wavelength at the resonance is approximately $\frac{1}{\sqrt{ME_0}}\sim10\ \ma{fm}$. We thus have well-separated length scales: $0.7\ \ma{fm}<2\ \ma{fm}<10\ \ma{fm}$, indicating the validity of EFT in describing the low-energy $\ma{\alpha}$-$\ma{\alpha}$ scattering. 

The nuclear EFT was first constructed and used to study the low-energy nucleon-nucleon scattering \cite{Kaplan:1996xu,Kaplan:1998we}. It has been extended to include Coulomb effects \cite{Kong:1999sf} and was used to study the $\alpha$-$\alpha$ scattering \cite{Higa:2008dn}. The $\alpha$-$\alpha$ nuclear potential has also been derived via a two-pion exchange in a chiral EFT and used to study the lifetime of $^8$Be \cite{Arriola:2007de}. In the pionless EFT framework, the $\ma{\alpha}$ particle is described by a scalar field $N$ with a mass $M$. The only interactions involved in the effective Lagrangian are contact interactions since the interaction length scale is comparable or smaller than the size of the $\ma{\alpha}$ particle. The effective Lagrangian is
\be
\ml{L}=N^{\dagger}\bigg(i\partial_t+\frac{\nabla^2}{2M}\bigg)N+\bigg[ C_0(N^{\dagger}N)^2 + C_2^{(1)}\Big(N^{\dagger} \overleftrightarrow{\nabla}  N\Big)^2 
+C_2^{(2)}\Big(i\overrightarrow{\nabla} (N^{\dagger}N)\Big)^2 +\cdots \bigg].
\ee
The four-point vertex with an incoming momentum ${\bf p}$ in the CM frame is
$-i\sum_{n=0}^{\infty}C_{2n}p^{2n}$,
where $C_{2n}$ is a linear combination of the Lagrangian parameters $C_{2n}^{(i)}$. In loop corrections, each vertex can still be written this way with ${\bf p}$ being the external line's momentum \cite{Beane:2000fi}.
To include the Coulomb effect, one can replace the ordinary derivative with the covariant derivative. Due to the non-relativistic nature of the Lagrangian, the magnetic interactions are suppressed by the large mass factor, and only the Coulomb interaction matters. The Coulomb modification on the nuclear scattering has been shown to be non-perturbative, which requires an infinite re-summation of Feynman diagrams \cite{Kong:1999sf}.
\begin{figure}
\centering
\includegraphics[width=0.55\linewidth]{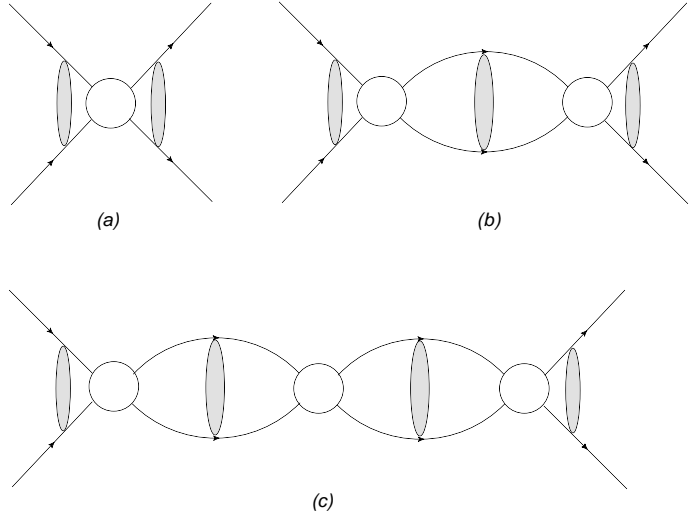}
\caption{First three Feynman diagrams contributing to the $\alpha$-$\alpha$ nuclear scattering amplitude with Coulomb corrections. The grey blob indicates the Coulomb Green's function given in Fig.~\ref{fig:coulomb green}.}
\label{fig:feynman}
\end{figure}
\begin{figure}
\centering
\includegraphics[width=0.7\linewidth]{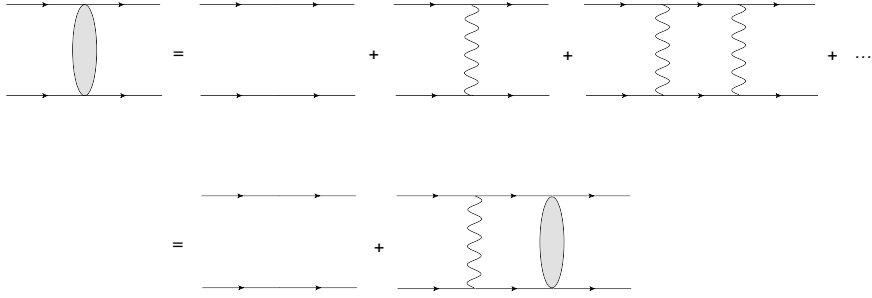}
\caption{Coulomb Green's function and its iterative calculation.}
\label{fig:coulomb green}
\end{figure}

\begin{figure}
\centering
\includegraphics[width=0.7in]{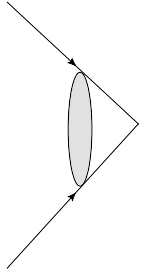}
\caption{Diagrammatic interpretation of $\chi_{\bf p}(0)$.}
\label{fig:chi}
\end{figure}

The $\ma{\alpha}$-$\ma{\alpha}$ scattering amplitude has two parts: one is purely from the Coulomb exchanges, the other derives from the nuclear interaction modified by the Coulomb repulsion. The latter is given by the summation of a geometric series of Feynman diagrams. The first three are shown in Fig.~\ref{fig:feynman}:
\be\label{Tmat}
T&=&-\frac{\big|\chi_{\bf p}(0)\big|^2e^{2i\sigma_0}}{(\sum_nC_{2n}p^{2n})^{-1}-G(E,0,0)},
\ee
where $\sigma_0$ is the phase shift caused by the Coulomb interaction only. Here, $G(E,{\bf r}',{\bf r})=\langle {\bf r}'|\hat{G}_C^{(+)}(E)|{\bf r}\rangle$ is the retarded Green's function in the spatial representation, which is also the two-line propagator connecting two vertices. The retarded Green's function is
\be
\hat{G}_C^{(+)}(E)=\frac{1}{E-\hat{H}_0-V_C+ i\epsilon},
\ee
where $\hat{H}_0={\bf \hat{p}}^2/M$. When $V_C=0$, this is the free Green's function $\hat{G}_0^{(+)}$. In the zero-temperature case $V_C=Z_1Z_2\alpha/r$. In a finite-temperature QED
plasma, effectively the internal photon becomes massive and the Coulomb repulsion is screened, becoming a Yukawa repulsion $V_C=(Z_1Z_2\alpha/r)e^{-m_Dr}$ where $m_D$ is the Debye mass depending on the plasma temperature $T$. The interacting Green's function can be calculated iteratively, corresponding to the diagrams shown in Fig.~\ref{fig:coulomb green}:
\be
\hat{G}_C^{(+)}=\hat{G}_0^{(+)}+\hat{G}_0^{(+)}V_C\hat{G}_C^{(+)}.
\ee
It is also the summation of all possible Coulomb exchanges of photons between the two internal scalar lines. Here, $\chi_{\bf p}(0)$ is the wave function at the origin for the Hamiltonian $H=H_0+V_C$ and can be represented by the incoming/outgoing diagram shown in Fig.~\ref{fig:chi}. In the zero-temperature limit, $|\chi_{\bf p}(0)|^2=C_{\eta}^2=\frac{2\pi\eta}{e^{2\pi\eta}-1}$ is the Sommerfeld factor where $\eta=\frac{Z_1Z_2\alpha M}{2p}$.

We will use Eq.~(\ref{Tmat}) in a manner that respects unitarity exactly. In the low energy domain, we expect that the higher-order terms in $p^2$ in the denominator contribute less, so  we expand the sum $(\sum_nC_{2n}p^{2n})^{-1}$ to order $p^4$ and obtain
\be
T&=&-\frac{\big|\chi_{\bf p}(0)\big|^2e^{2i\sigma_0}}{{\frac{1}{C_0}-\frac{C_2}{C_0^2}p^2+\big( \frac{C_2^2}{C_0^3}-\frac{C_4}{C_0^2} \big)p^4}-G(E,0,0)},
\ee
which corresponds to the effective range expansion truncated to next-to-next-to-leading order, while preserving exact unitarity. Expanding the effective range has been shown to be better than expanding the scattering amplitude in powers of $p$ at reproducing the phase shift \cite{Higa:2008dn}. The scattering amplitude  is related to the phase shift $\delta_0$ of the nuclear interaction modified by the Coulomb repulsion via
\be
T&=&\frac{4\pi}{M}\frac{e^{2i\sigma_0}}{p\cot\delta_0 -ip}.
\ee
Comparing the two expressions leads to a formula for the Coulomb modified nuclear phase shift in terms of the EFT parameters:
\be
p\cot\delta_0 -ip &=&-\frac{4\pi/M}{\big|\chi_{\bf p}(0)\big|^2}\bigg( \frac{1}{C_0}-\frac{C_2}{C_0^2}p^2+\Big( \frac{C_2^2}{C_0^3}-\frac{C_4}{C_0^2} \Big) p^4 -G(E,0,0)  \bigg).
\ee
The unscreened Coulomb Green's function in the dimensional regularization and MS renormalization scheme is given by \cite{Kong:1999sf}
\be
G(E,0,0; m_D=0) &=& \frac{Z_1Z_2\alpha M^2}{4\pi}\bigg( \frac{1}{\epsilon}-H(\eta)+\ln{\frac{\mu\sqrt{\pi}}{Z_1Z_2\alpha M}}+1-\frac{3}{2}\gamma \bigg),
\ee
where $\mu$ is the renormalization scale and
\be
H(\eta)&=&\psi(i\eta)+\frac{1}{2i\eta}-\ln{i\eta}=\Re\psi(1+i\eta)+\frac{iC_{\eta}^2}{2\eta}-\ln{\eta},
\ee 
in which $\psi(x)$ denotes the Digamma function.
The screened Yukawa Green's function can be calculated by numerically solving a quantum mechanical scattering problem with a delta-function and Yukawa potential \cite{Kaplan:1996xu}. The nuclear phase shift is computed by studying and matching the asymptotic behaviors of the regular solution $J_E(r)$ and the irregular solution $K_E^{\lambda}(r)$. $\lambda$ is determined by the boundary condition and the final expression of the Green's function is independent of it. Both solutions are free waves asymptotically because the Yukawa potential decays exponentially,
\be
J_E(r)&\xrightarrow{r\rightarrow\infty}&y\frac{e^{ipr}}{pr}+y^*\frac{e^{-ipr}}{pr},\\
K_E^{\lambda}(r)&\xrightarrow{r\rightarrow\infty}&z\frac{e^{ipr}}{pr}+z^*\frac{e^{-ipr}}{pr}.
\ee
The result for the Green's function is given by \cite{Kaplan:1996xu}
\be
\label{eq:md_green}
G(E,0,0;m_D\ne0) &=& \frac{z^*}{y^*}-\lim_{r\rightarrow0}K_E^{\lambda}(r).
\ee
The second term is divergent, and this divergence can be absorbed into the definition of the constant $C_0$.  The numerical procedure also gives the wave function at the origin for the screened case $\chi_{\bf p}(0)=-\frac{1}{2iy^*}$ \cite{Kaplan:1996xu}.

We want to use the zero-temperature experimental data to fit the EFT parameters and then apply them to the finite-temperature case. The singularities in the Coulomb ($m_D=0$) and the Yukawa ($m_D\ne0$) Green's functions are the same. One can explicitly show this by computing the no-photon and one-photon exchange diagrams as they are the only divergent diagrams. However, the calculation of the Green's function in the $m_D=0$ limit can be done analytically using dimensional regularization as a regulator, while
for $m_D \neq 0$, this is done numerically with the regulator defined by Eq.~(\ref{eq:md_green}). Therefore the renormalized  coupling $C_0$ is not the same in the two calculations. We expect there to be an additive constant relating the coupling constants in the two renormalization schemes. (For a perturbative calculation of this constant, see Ref.~\cite{Kaplan:1996xu}.) In our calculation, we fix this constant numerically by demanding that as $m_D \to 0$ the numerically computed Green's function coincide with the 
renormalized Coulomb Green's function. 

To proceed, we absorb the $\epsilon$-pole and all the constants in the Coulomb Green's function into the parameter $C_0$ such that the phase shift in the zero-temperature case is given by (noticing that the imaginary parts on both sides cancel)
\be
\label{eq:phaseshift}
\cot\delta_0=\frac{1}{pC_{\eta}^2}\bigg(  -\frac{4\pi}{M}\frac{1}{C_0} + \frac{4\pi}{M}\frac{C_2}{C^2_0}p^2-\frac{4\pi}{M}\Big( \frac{C_2^2}{C_0^3}-\frac{C_4}{C_0^2} \Big)p^4-Z_1Z_2\alpha M \Re H(\eta)  \bigg).
\ee
Comparing with the standard effective range expansion of the Coulomb wave function phase shift,
\be
C_{\eta}^2p\cot\delta_0+Z_1Z_2\alpha M \Re H(\eta)=-\frac{1}{a}+\frac{1}{2}r_0p^2-\frac{1}{4}P_0p^4+\cdots
\ee
where $a$ is the scattering length, $r_0$ the effective range and $P_0$ the shape parameter, we find
\be
-\frac{1}{a}&=&-\frac{4\pi}{M}\frac{1}{C_0}\equiv A,\\
\frac{1}{2}r_0&=&\frac{4\pi}{M}\frac{C_2}{C_0^2}\equiv B,\\
-\frac{1}{4}P_0&=&-\frac{4\pi}{M}\Big(  \frac{C_2^2}{C_0^3}-\frac{C_4}{C_0^2} \Big)\equiv C.
\ee

\begin{figure}
\centering
\includegraphics[width=0.5\linewidth]{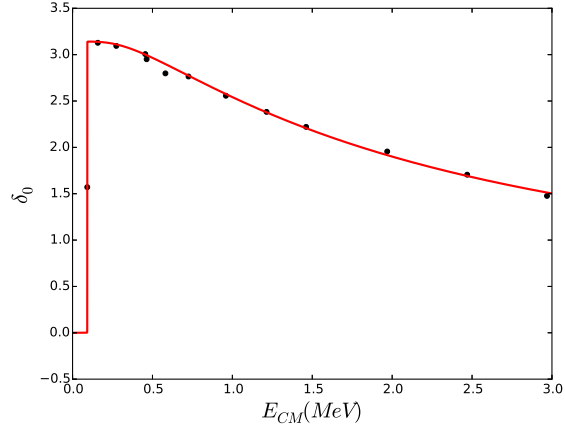}
\caption{Phase shift caused by the nuclear interaction modified by the unscreened Coulomb repulsion up to $E_{CM}$=3 MeV. Black dots are the experimental data from Ref.~\cite{Russell:1956zz}.}
\label{fig:phase shift}
\end{figure}

Then we fit these three parameters by using the experimentally determined resonance energy, width and the phase shift up to $E_{CM}=3\ \ma{MeV}$. The resonance at $E_{CM}\equiv E_0$ corresponds to a $s$-wave phase shift $\delta_0=\frac{\pi}{2}$, i.e., $\cot\delta_0=0$. The width of the resonance $\Gamma$ is given by 
\be
\label{eq:width}
\frac{d\cot\delta_0(E)}{dE}\Big|_{E=E_0}\equiv-\frac{2}{\Gamma}.
\ee
We use Eq.~(\ref{eq:phaseshift}) and Eq.~(\ref{eq:width}) to calculate the resonance energy, width and phase shift and apply a least square fit of the parameters. The best fit result is shown in Table~\ref{table:1}. The best fitted resonance energy and width are summarized in Table~\ref{table:2} with the experimental data. The calculated phase shift is plotted in Fig.~\ref{fig:phase shift} along with the experimental measurements from Ref.~\cite{Russell:1956zz}. The agreement with experimental data is excellent. Our values for the extracted scattering length, effective range, and shape parameter are consistent with a similar fit in Ref.~\cite{Higa:2008dn}. Our numerical values differ slightly because we fit up to $E_{\rm CM} = 3$ MeV, while Ref.~\cite{Higa:2008dn} fits up to $E_{\rm lab}=3$ MeV, which corresponds to $E_{\rm CM} =1.5$ MeV.
 
\begin{table*}[htdp]
\caption{\label{table:1}Best fit parameters}
\begin{center}
\begin{tabular}{|c|c|c|c|}
\hline
Parameter & $a\ (10^3\ \ma{fm})$ & $r_0\ (\ma{fm})$ & $P_0\ (\ma{fm}^3)$\\
\hline
Best fit value (accurate to $10^{-3}$) &  -2.029 & 1.104 &  -1.824  \\
\hline
\end{tabular}
\end{center}
\end{table*}

\begin{table*}[htdp]
\caption{\label{table:2}Best fitted resonance energy and width}
\begin{center}
\begin{tabular}{|c|c|c|}
\hline
Physical quantity & Resonance energy $E_0$ (keV) & Width $\Gamma$ (eV)\\
\hline
Best fitted value (accurate to $10^{-3}$) &   91.838   &   5.715   \\
\hline
Experimental value \cite{Wustenbeckeretal.1992} & 91.84 $\pm$ 0.04   &  5.57 $\pm$ 0.25 \\
\hline
\end{tabular}
\end{center}
\end{table*}

\begin{figure}[htdp]
\centering
\includegraphics[width=0.47\linewidth]{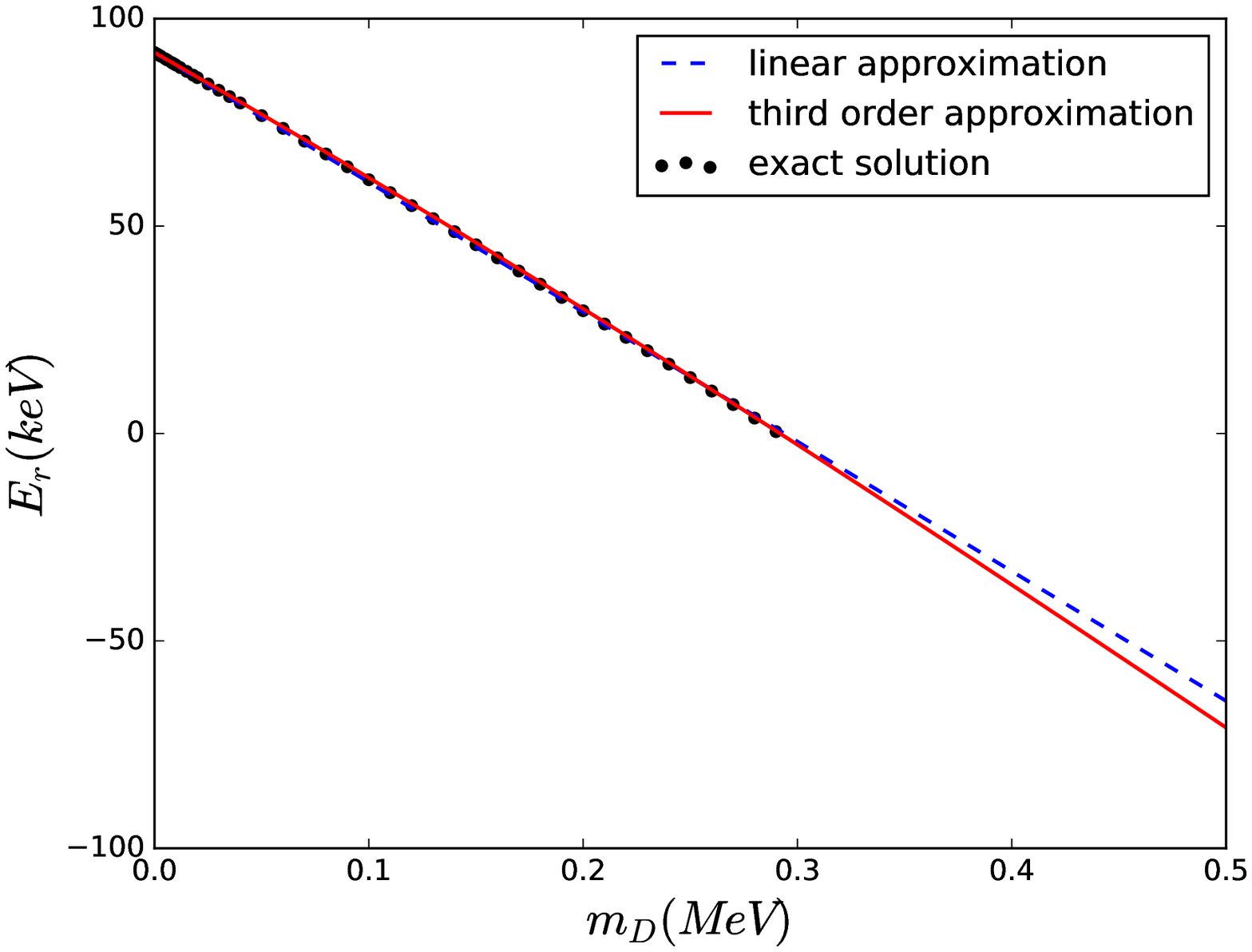}
\includegraphics[width=0.47\linewidth]{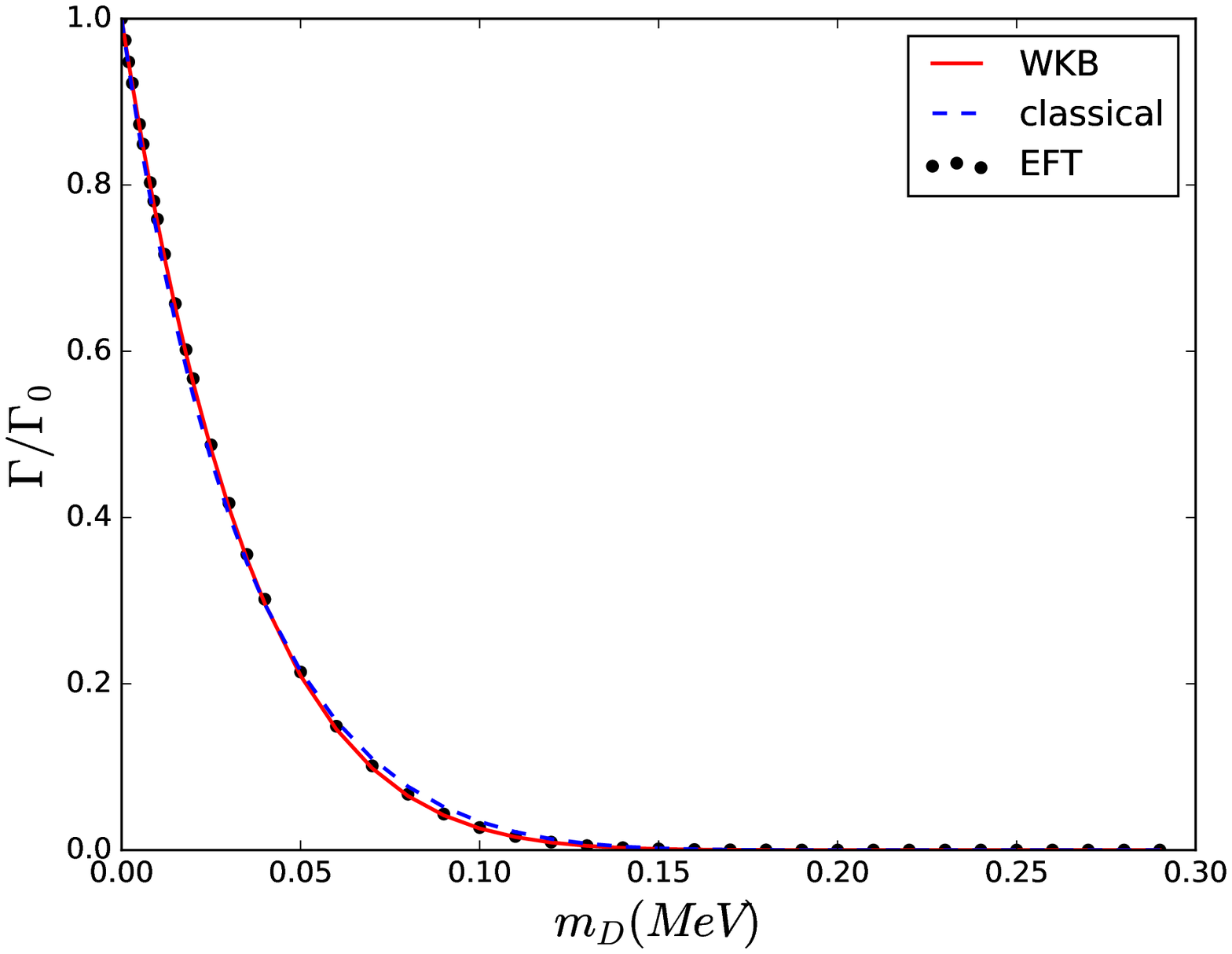}
\caption{Finite temperature results for the resonance energy and the width.}
\label{fig:finite_reso_width}
\end{figure}

Now we move on to the screened case. The nuclear phase shift is
\be
\label{eq:phaseshiftyukawa}
\cot\delta_0=\frac{1}{p|\chi_{\bf p}(0)|^2}\bigg( A+Bp^2+Cp^4+\frac{4\pi}{M}\Re\Big(  \frac{z^*}{y^*}+C_{renor} \Big)  \bigg),
\ee
where we have written the Yukawa Green's function as $\frac{z^*}{y^*}+C_{renor}$. Again the imaginary parts cancel. As discussed above, we fix $C_{renor}$ numerically by calculating the Yukawa Green's function at $m_D=0.001$, $0.002$, $0.003$, $0.004$, $0.005$ MeV. We find that the dependence on $m_D$ is almost linear when $m_D\leq0.005$ MeV, which can also be shown analytically by expanding the Yukawa Green's function in terms of $m_D$. Therefore, we interpolate the Yukawa Green's function linearly towards $m_D=0$. By matching the real part of the Yukawa with that of the renormalized Coulomb Green's function $-Z_1Z_2\alpha M^2 \Re H(\eta)/4\pi$ for $m_D=0$, we find $C_{renor}\approx 156321.8617$. Then we use Eq.~(\ref{eq:phaseshiftyukawa}) and Eq.~(\ref{eq:width}) to compute the resonance energy and the width. The result is shown in Fig.~\ref{fig:finite_reso_width}.

Along with the EFT calculation results, we also plot the first order approximation and the third order approximation of the resonance energy, which can be extended to the bound state formation energy region where the EFT framework does not have an explicit analytic continuation because the Yukawa Green's function has no analytic expression. The resonance corresponds to $\cot\delta_0=0$, i.e.,
\be
\label{eq:solve_reso}
A+Bp^2+Cp^4+\frac{4\pi}{M}\Re\big[G({E,0,0;m_D})\big]=0.
\ee
For simplicity we will write the Green's function as $G({E,0,0;m_D})=G(E,m_D)$. 
Using $p^2=ME$ and the fact that $E_0$ is a solution to Eq.~(\ref{eq:solve_reso}) when $m_D=0$, one can expand the Yukawa Green's function around $E_0$ and $m_D=0$ to obtain
\be
\label{eq:3rd}
0=BM(E-E_0)+CM^2(E^2-E_0^2)+\frac{4\pi}{M}\Re\bigg[  \frac{\partial G}{\partial E}\Big|_{(E_0,0)}(E-E_0)   +\\
\nonumber
 \frac{1}{2}\frac{\partial^2 G}{\partial E^2}\Big|_{(E_0,0)}(E-E_0) ^2  +\frac{1}{6}\frac{\partial^3 G}{\partial E^3}\Big|_{(E_0,0)}(E-E_0)^3 + 
 \frac{\partial G}{\partial m_D}\Big|_{(E_0,0)}m_D +\cdots \bigg]
\ee
where the $\cdots$ indicates higher-order corrections. Since near $m_D=0$ the Green's function changes almost linearly, we only expand with respect to $m_D$ to the first order and numerically $\frac{4\pi}{M}\Re\big(\frac{\partial G}{\partial m_D}\big|_{(E_0,0)}\big)\approx -0.3334$. The renormalized Coulomb Green's function $4\pi G(E,0)/M=-Z_1Z_2\alpha MH(\eta)$ depends on the energy and thus can be used to estimate the partial derivatives with respect to $E$. The third order approximation in the plot corresponds to solving Eq.~(\ref{eq:3rd}) for the resonance energy. If we make the first order approximation, we will have a much simpler expression.
Since we have omitted higher order terms, the Taylor series theorem tells us that the best value of $\frac{\partial G}{\partial E}$ to use is that at some $\xi\in[0, E_0]$. If we set $\xi=E_0/2$, we will have
\be
E = E_0-\frac{\frac{4\pi}{M}\Re\Big(\frac{\partial G}{\partial m_D}\big|_{(E_0,0)}\Big)m_D}{BM+\frac{4\pi}{M}\Re\big(\frac{\partial G}{\partial E}\big|_{(\frac{E_0}{2},0)}\big)},
\ee
which corresponds to the linear approximation in Fig.~\ref{fig:finite_reso_width}. This best explains the almost linear behavior of the resonance energy as a function of the screening mass.

Now we move on to the bound state formation, which corresponds to the pole of the scattering amplitude in the negative energy region. When $E>0$, $G(E,m_D)$ has both real and imaginary parts. When analytically continuing $p$ to $ip$, the energy becomes negative. At the same time the imaginary part becomes real but it turns out to be negligible. So Eq.~(\ref{eq:solve_reso}) and its approximation Eq.~(\ref{eq:3rd}) are still valid for solving bound state energy. The result is the negative energy part in Fig.~\ref{fig:finite_reso_width}.

For the width, the Gamow classical expression and the WKB method are also plotted for comparison. The classical turning point of Coulomb repulsion is given by setting $\frac{Z_1Z_2\alpha}{r_0}$ equal to the resonance energy $E_r=\frac{p_r^2}{M}$, which gives $r_0=\frac{Z_1Z_2\alpha}{E_r}$. The barrier tunneling probability is given by the Gamow factor $\Gamma=e^{-\pi r/\lambda_r}$,
where $\lambda_r=\frac{1}{p_r}=\frac{1}{\sqrt{ME_r}}$. Since we are considering the Yukawa repulsion, one estimate is to use $r=r_0e^{-m_Dr_0}$ and obtain 
\be
\Gamma=\exp{-\sqrt{\frac{(\pi Z_1Z_2\alpha)^2M}{E_r}}\big(e^{-m_DZ_1Z_2\alpha/E_r}\big)},
\ee
which corresponds to the classical curve in Fig.~\ref{fig:finite_reso_width}.
A better estimate is given by the WKB method whereby $\Gamma\propto e^{-2\phi}$ and $\phi=\int_a^b\sqrt{M\big(V_C(x)-E\big)}dx$ where $V_C(x)$ is the unscreened/screened Coulomb potential. The lower and upper limits
$a$ and $b$ are the two classical turning points of $V_C(x)-E$. Here $a=0$ since we have a delta potential attraction to describe the nuclear potential and $b$ is obtained by solving $V_C(b)=E_r$.

We see that due to the QED plasma screening the resonance energy drops while the lifetime against the spontaneous decay into two $\alpha$ particles increases. The resonance becomes a bound state when $m_D\gtrsim 0.3$ MeV. This threshold behavior can be further tested experimentally in the following way: prepare a $^7$Li metal target surrounded by hydrogen gas and shine high-intensity lasers onto the system to heat them up. Locally the $^7$Li and hydrogen atoms become completely ionized and an $e^-e^+$ plasma is formed. The Debye mass, which depends on the plasma temperature, can be tuned by changing the laser intensity. The $^7$Li and proton scattering produces $^8$Be (through the decay of excited $^8$Be states) that decays to two $\alpha$ particles in low temperature. By searching for the two $\alpha$ particles event at different laser intensities, one can tell under what temperature or Debye mass the resonance becomes a bound state. The experiment design requires more careful feasibility considerations.

\begin{figure}
\centering
\includegraphics[width=0.5\linewidth]{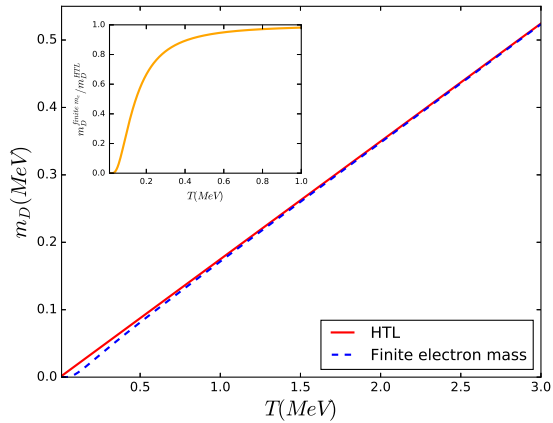}
\caption{Debye mass in the HTL approximation and with $m_e$ retained.}
\label{fig:md}
\end{figure}

For practical purposes, an explicit formula relating the Debye mass and the plasma temperature is needed. For the relativistic $e^-e^+$ plasma in the early universe, one might use the Hard Thermal Loop (HTL) result $m^{HTL}_D=\sqrt{\frac{4\pi\alpha}{3}}T$ \cite{Bellac:2011kqa}. The HTL assumes $T\gg m_e$ and set the electron mass $m_e$ to be zero. But the starting temperature of nuclei formations $T\sim1$ MeV is comparable with $m_e$. To include the finite electron mass correction, we compute the photon self-energy to one-loop in the imaginary time formalism of thermal field theory
\be
\Pi_{\mu\nu}(Q)=e^2\int\frac{d^4K}{(2\pi)^4}\Tr{\gamma_{\mu}\Big(\slashed{K}-m_e\Big)\gamma_{\nu}\Big((\slashed{K}-\slashed{Q})-m_e\Big)} \tilde{\Delta}(K) \tilde{\Delta}(K-Q),
\ee
where in the Euclidean space the internal fermion (electron or positron) momenta $K=(k_4,\bs{k})=(-\omega_n,\bs{k})$ and $\omega_n=(2n+1)\pi T$ are the Matsubara frequencies. $Q=(q_0, \bs{q})$ is the external photon momentum. The denominator of the fermion propagator is given by 
\be
\tilde{\Delta}(K)=\frac{1}{\omega_n^2+E_k^2},
\ee
with $E_k=\sqrt{\bs{k}^2+m_e^2}$. The momentum integral includes a summation over the Matsubara frequencies
\be
\int\frac{d^4K}{(2\pi)^4}\equiv T\sum_n\int\frac{d^3k}{(2\pi)^3}.
\ee
For the Debye mass in the static screening, we compute $\Pi_{00}$ with $q_0=0$ and then take the limit $\bs{q}\rightarrow0$. The result is given by
\be
m_D^2&=&-\Pi_{00}(q_0=0,\bs{q}\rightarrow0)= -4e^2\int\frac{d^3k}{(2\pi)^3}\frac{\partial\, n(E_k)}{\partial E_k}\\
&=&\frac{8\alpha}{\pi T}\int_0^{\infty}k^2dk\frac{e^{\beta E_k}}{(e^{\beta E_k}+1)^2}. \nn
\ee
The HTL and the finite $m_e$ results and their ratios are plotted in Fig.~\ref{fig:md}. From the ratio subplot, we conclude that the HTL result overestimates the Debye mass when $T<1$ MeV.

The modifications on the $^8$Be resonance energy and lifetime due to the plasma screening may be relevant to the ``lithium problem'' in the Big Bang Nucleosynthesis (BBN). The problem is a serious discrepancy between the theory and the experiment concerning the primordial abundance of $^7\ma{Li}$. With the precision measurements of the cosmic microwave background (CMB) from WMAP \cite{Hinshaw:2012aka} and Planck \cite{Ade:2015xua}, the BBN calculation is improved with the baryon-photon number density ratio $\eta_b$ as an input parameter \cite{Cyburt:2003fe}. The current $^7\ma{Li}$ prediction is larger than the experimental value by $4-5$ standard deviations \cite{Cyburt:2008kw,Cyburt:2015mya}. Many previous studies have tried to resolve the problem (a good review is given in Ref.~\cite{Fields:2011zzb}). One perspective focuses on a more accurate determination of nuclear reaction rates. In particular, the plasma screening effects on non-resonant nuclear reaction rates have been shown to be negligible \cite{1997ApJ...488..507I,Wang:2010px}. In principle, the resonant states created in the $A=7$ element destructions could provide a solution \cite{Cyburt:2009cf,Chakraborty:2010zj} but a recent study suggests this is also insignificant \cite{Famiano:2016hhs}.

An important alternative reaction related to the $^7$Li abundance is the charge exchange reaction $^7$Be(n, p)$^7$Li, where an excited state of $^8$Be with $(J^P, I)=(2^-,0)$ exists approximately $0.01$ MeV above the $^7$Be+n threshold, which can decay to the ground state \cite{Tilley:2004zz}. A coupled-channel EFT has been derived to study this reaction in vacuum \cite{Lensky:2011he}. Our calculation shows the improved stability of the ground state in the plasma, but is incomplete in the sense that only the static screening effect is considered. Therefore it would be interesting to include the dynamic screening in the calculation of the $^8$Be system. One key dynamic screening effect is the thermal width caused by collisions with electrons/positrons in the plasma. We will study this in future work.

Our results also have applications in other nuclear fusion processes such as the stellar nucleosynthesis in the helium burning stage. Inside the star core, due to the gravitational collapse typical temperatures can be several hundred eV or higher. Atoms are completely ionized, resulting in a non-relativistic electron plasma with a finite density. Under a given temperature and electron density, the Debye mass can be calculated assuming that the plasma consists of electrons only. Our calculations can then be directly applied. Due to the plasma screening, the $^4$He burning product $^8$Be lives longer and has a higher chance to collide with another $^4$He to form $^{12}$C, which is stable. To fully understand and simulate the reaction chain, one has to include the modification on the lifetime caused by the plasma effect. In this sense, our results could deepen our understanding of stellar evolution.

\begin{acknowledgments}
We thank Sean Fleming, Robert Pisarski and Johann Rafelski for very helpful discussions. B.M. and X.Y. are supported by U.S. Department of Energy research grant DE-FG02-05ER41367,
T.M. is supported by U.S. Department of Energy research grant DE-FG02-05ER41368. T.M. and X.Y. would like to thank the theory group at Brookhaven National Laboratory  for their hospitality during the completion of this work.

\end{acknowledgments}


\begin{thebibliography}{99}

\bibitem{Wustenbeckeretal.1992}
 S.~W\"ustenbecker, H.~W.~Becker, H.~Ebbing, W.~H.~Schulte, M.~Berheide, M.~Buschmann, C.~Rolfs, G.~E.~Mitchell, and J.~S.~Schweitzer,
 Zeitschrift\ f\"ur\ Physik\ A\ Hadrons\ and\ Nuclei {\bf 344}, 205 (1992).
 
\bibitem{Russell:1956zz} 
  J.~L.~Russell, G.~C.~Phillips and C.~W.~Reich,
  Phys.\ Rev.\  {\bf 104}, 135 (1956).

 
\bibitem{Kaplan:1996xu} 
  D.~B.~Kaplan, M.~J.~Savage and M.~B.~Wise,
  Nucl.\ Phys.\ B {\bf 478}, 629 (1996)
  [nucl-th/9605002].
  
\bibitem{Kaplan:1998we} 
  D.~B.~Kaplan, M.~J.~Savage and M.~B.~Wise,
  Nucl.\ Phys.\ B {\bf 534}, 329 (1998)
  [nucl-th/9802075].
  
\bibitem{Kong:1999sf} 
  X.~Kong and F.~Ravndal,
  Nucl.\ Phys.\ A {\bf 665}, 137 (2000)
  [hep-ph/9903523].
  

\bibitem{Higa:2008dn} 
  R.~Higa, H.-W.~Hammer and U.~van Kolck,
  Nucl.\ Phys.\ A {\bf 809}, 171 (2008)
  [arXiv:0802.3426 [nucl-th]].


\bibitem{Arriola:2007de} 
  E.~Ruiz Arriola,
  arXiv:0709.4134 [nucl-th].


\bibitem{Beane:2000fi} 
  S.~R.~Beane and M.~J.~Savage,
  Nucl.\ Phys.\ A {\bf 694}, 511 (2001)
  [nucl-th/0011067].


\bibitem{Bellac:2011kqa}
 M.~L.~Bellac,
 {\it Thermal Field Theory} (Cambridge University Press, 2011).
 

  
\bibitem{Hinshaw:2012aka} 
  G.~Hinshaw {\it et al.} [WMAP Collaboration],
  Astrophys.\ J.\ Suppl.\  {\bf 208}, 19 (2013)
  [arXiv:1212.5226 [astro-ph.CO]].
  
\bibitem{Ade:2015xua} 
  P.~A.~R.~Ade {\it et al.} [Planck Collaboration],
  arXiv:1502.01589 [astro-ph.CO].
  
\bibitem{Cyburt:2003fe} 
  R.~H.~Cyburt, B.~D.~Fields and K.~A.~Olive,
  Phys.\ Lett.\ B {\bf 567}, 227 (2003)
  [astro-ph/0302431].
  
\bibitem{Cyburt:2008kw} 
  R.~H.~Cyburt, B.~D.~Fields and K.~A.~Olive,
  JCAP {\bf 0811}, 012 (2008)
  [arXiv:0808.2818 [astro-ph]].
  
\bibitem{Cyburt:2015mya} 
  R.~H.~Cyburt, B.~D.~Fields, K.~A.~Olive and T.~H.~Yeh,
  Rev.\ Mod.\ Phys.\  {\bf 88}, 015004 (2016)
  [arXiv:1505.01076 [astro-ph.CO]].

\bibitem{Fields:2011zzb} 
  B.~D.~Fields,
  Ann.\ Rev.\ Nucl.\ Part.\ Sci.\  {\bf 61}, 47 (2011)
  [arXiv:1203.3551 [astro-ph.CO]].


\bibitem{1997ApJ...488..507I}
N.~Itoh, A.~Nishikawa, S.~Nozawa, and Y.~Kohyama,
 Astrophys.\ J.\ {\bf 488}, 507 (1997).
 
 
\bibitem{Wang:2010px} 
  B.~Wang, C.~A.~Bertulani and A.~B.~Balantekin,
  Phys.\ Rev.\ C {\bf 83}, 018801 (2011)
  [arXiv:1010.1565 [astro-ph.CO]].

\bibitem{Cyburt:2009cf} 
  R.~H.~Cyburt and M.~Pospelov,
  Int.\ J.\ Mod.\ Phys.\ E {\bf 21}, 1250004 (2012)
  [arXiv:0906.4373 [astro-ph.CO]].

\bibitem{Chakraborty:2010zj} 
  N.~Chakraborty, B.~D.~Fields and K.~A.~Olive,
  Phys.\ Rev.\ D {\bf 83}, 063006 (2011)
  [arXiv:1011.0722 [astro-ph.CO]].
  
\bibitem{Famiano:2016hhs} 
  M.~A.~Famiano, A.~B.~Balantekin and T.~Kajino,
  Phys.\ Rev.\ C {\bf 93}, 045804 (2016)
  [arXiv:1603.03137 [astro-ph.CO]].

\bibitem{Tilley:2004zz} 
  D.~R.~Tilley, J.~H.~Kelley, J.~L.~Godwin, D.~J.~Millener, J.~E.~Purcell, C.~G.~Sheu and H.~R.~Weller,
  Nucl.\ Phys.\ A {\bf 745}, 155 (2004).

\bibitem{Lensky:2011he} 
  V.~Lensky and M.~C.~Birse,
  Eur.\ Phys.\ J.\ A {\bf 47}, 142 (2011)
  [arXiv:1109.2797 [nucl-th]].

\end{thebibliography}
\end{document}